# Wideband tunable microwave signal generation in a silicon-based optoelectronic oscillator

**Phuong T.Do**[1,*], **Carlos Alonso-Ramos**[2], **Xavier Le Roux**[2], **Isabelle Ledoux**[1], **Bernard Journet**[1], **and Eric Cassan**[2]

[1]LPQM (CNRS UMR-8537,École normale supérieure Paris-Saclay, Université Paris-Saclay, 61 avenue du Président Wilson, 94235 Cachan Cedex, France
[2]C2N (CNRS UMR-9001), Univ. Paris-Sud, Université Paris-Saclay, 10 Boulevard Thomas Gobert, 91120 Palaiseau, France
[*]tdo01@ens-paris-saclay.fr

## ABSTRACT

Si photonics has an immense potential for the development of compact and low-loss opto-electronic oscillators (OEO), with applications in radar and wireless communications. However, current Si OEO have shown a limited performance. Si OEO relying on direct conversion of intensity modulated signals into the microwave domain yield a limited tunability. Wider tunability has been shown by indirect phase-modulation to intensity-modulation conversion, requiring precise control of the phase-modulation. Here, we propose a new approach enabling Si OEOs with wide tunability and direct intensity-modulation to microwave conversion. The microwave signal is created by the beating between an optical source and single sideband modulation signal, selected by an add-drop ring resonator working as an optical bandpass filter. The tunability is achieved by changing the wavelength spacing between the optical source and resonance peak of the resonator. Based on this concept, we experimentally demonstrate microwave signal generation between 6 GHz and 18 GHz, the widest range for a Si-based OEO. Moreover, preliminary results indicate that the proposed Si OEO provides precise refractive index monitoring, with a sensitivity of 94350 GHz/RIU and a potential limit of detection of only $10^{-8}$ RIU, opening a new route for the implementation of high-performance Si photonic sensors.

## Introduction

The generation of broadband and low noise microwave and millimeter wave signals is important for many applications, including among others, radars, wireless communications, optical signal processing, warfare systems, and modern instrumentation[1,2,3]. Among the different approaches to generate microwave and millimeter signals, the optoelectronic oscillator (OEO) is a particularly interesting solution due to its capability to provide, by direct synthesis, spectrally pure and wideband tunable signals[4,5]. A classical OEO has a fundamentally multi-mode behavior[6], with mode spacing associated with the km-long optical fiber delay lines used inside the closed-loop system. To select the desired oscillation mode, a microwave filter with high quality factor ($Q_{RF}$) is typically included inside the closed path[2,6]. To achieve variable frequency generation, this microwave filter needs to be tunable. However, a microwave filter with high $Q_{RF}$ and wide frequency tunability is practically hard to realize, especially for high operation frequencies[7]. In contrast, microwave photonic (MWP) are a promising alternative solution to overcome this limitation, allowing reconfigurable microwave signal generation in OEO with a wide tuning range[2,7,8]. In addition, the progress of integrated microwave photonics (IMWP)[9] provides now a solid framework for the full integration of an OEO. Several efforts in this direction have been demonstrated recently[10,11,12]. M. Merklein et al., in[10] demonstrated ultrawide frequency tunable signals up to 40 GHz by using OEO based on stimulated Brillouin scattering (SBS). However, the system explored therein is complicated as it requires harnessing light-sound interactions on chip, based on non-standard chalcogenide materials and the use of two lasers. In[11], an integrated optoelectronic oscillator based on InP was investigated, but the reported frequency tunability range was limited to only 20 MHz. On the other hand, the silicon on insulator (SOI) technology has been identified as a promising solution to implement ultra-compact and low-cost OEO, which could be fabricated using already existing large volume fabrication facilities. The unique potential of Si to integrate photonic and electronic functionalities within a single chip, together with the availability of high-performance key building blocks, e.g. all-Si modulators and Ge on Si photodetectors[13,14],[15], make Si an ideal candidate for the development of high-performance OEOs. However, the scarce demonstrations of Si-based OEOs showed a limited performance in terms of tunability. Direct conversion of intensity-modulated signals into the microwave domain has been shown based on quadratic detection of two successive transmission lines in the drop-port of the ring[16]. The microwave frequency is determined by the free-spectral-range (FSR) of the ring, limiting its tunability. In addition, microwave signal generation requires few-millimeters long ring resonators, which are difficult to implement. Microwave generation has also been demonstrated in Si-based OEO, implementing indirect phase-modulation to intensity-modulation conversion, relying on a notch filtering provided by a micro-disk operating in an all-pass configuration. This approach requires precise control of the phase modulation, and provided a limited tunability range between 3 and 6.8 GHz[12]. Here, we propose a new approach for the implementation of Si OEO that allows for wideband tunability in the microwave signal generation, based on a direct intensity-modulation to microwave conversion. As schematically shown in Fig.1a, the laser source is split in two paths. One path comprises an intensity modulator and an add-drop ring resonator (RR). The other path goes directly to the photodetector. The oscillation signal is created by the direct translation of the intensity modulation into the microwave domain, provided by the beating between the optical source (direct path) and one of the sideband lobes generated by the intensity modulator (path with intensity modulator and RR). This sideband lobe is selected by one transmission line of the silicon add-drop RR, that serves as optical bandpass filter. The frequency of the generated microwave signal is determined by the wavelength separation between the laser source and the resonance of the RR. By using only one of the transmission lines of the RR, we substantially relax the requirements on the free-spectral-range of the ring, while providing flexible tuning. We experimentally show that by tuning the wavelength of the source, the microwave frequency generated by the OEO can be tuned between 5.9 GHz and 18.2 GHz. This is, to the best of our knowledge, the widest tunability range reported for a Si-based OEO. A phase noise near -110 dBc/Hz at the offset frequency of 1 MHz, comparable with state-of-the-art photonic OEO[11,12], is measured for different oscillation frequencies along the 12 GHz tuning range. Concurrently, the proposed OEO performs a precise translation of the laser-to-RR wavelength separation into the microwave domain, where it can be precisely measured. Then, if the laser wavelength is fixed, monitoring of the microwave frequency shifts provides accurate information of the variations in the resonance wavelength of the RR, which can be related to variations in the refractive index. This way, by exploiting the improved spectral resolution in the microwave domain, the proposed OEO can also serve as a high-performance refractive index sensor. Preliminary experimental results show a sensitivity of 94350 GHz/RIU, i.e. a 40-fold improvement compared to previously reported microwave-photonic silicon refractive index sensors[17]. Based on the measured phase noise, we estimated a remarkably low achievable limit of detection (LOD) of only $10^{-8}$ RIU. These results illustrate the potential of this approach for the implementation of high-performance Si sensors, e.g. for lab-on-a-chip biosensing applications[18].

## Results

### Principle of operation

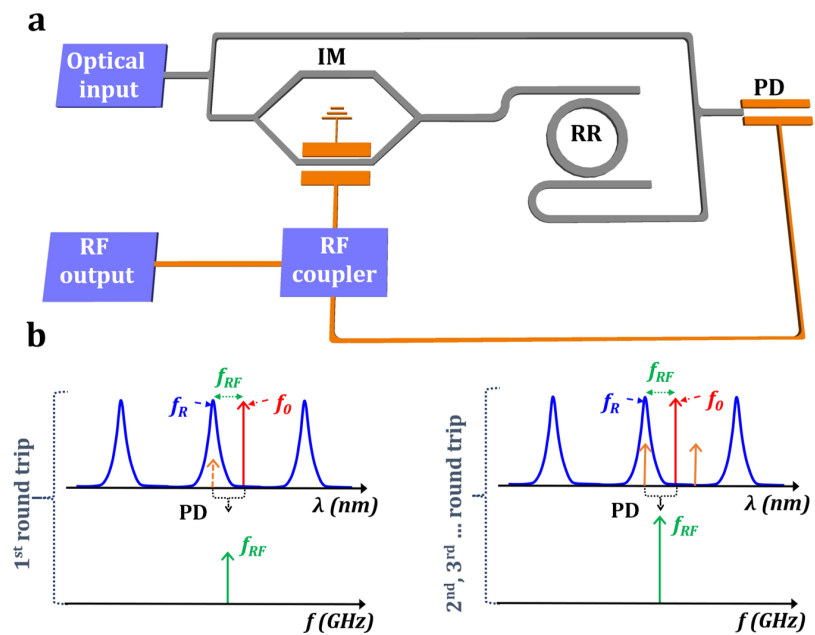

**Figure 1.** a Schematic of the proposed OEO structure and **b** Principle of operation of the proposed tunable OEO. In **a**, IM: Intensity modulator, RR: Ring resonator and PD: Photo-detector. In **b**, the red curve corresponds to the optical carrier (or laser source frequency); the orange curve illustrates the sideband lobes of the modulated signal, the blue curve indicates the optical transfer function of the RR and the green one represents the generated RF frequency $f_{RF}$.

In the proposed tunable OEO configuration, shown in Fig.1, the optical signal coming from the laser light source (frequency $f_0$) is separated into two arms. One is connected directly to the photodetector (PD), while the other feeds an intensity modulator (IM) followed by a silicon ring resonator (RR) in add-drop configuration. In this scheme, the input signal of the PD always comprises a part of the un-modulated laser light beam. At the initial stage, the modulator output signal grows, just seeded by white noise existing inside the loop. If one modulation output signal can go through the optical transfer function of the resonator at frequency $f_R$, this signal can then be combined with the optical carrier ($f_0$) at either its left or right sides to generate a beating of frequency $f_b$ at the input of the PD. If the distance between the optical carrier ($f_0$) and the signal at $f_R$ falls within the working range of the loop, the generated beating signal can be converted as a RF frequency $f_{RF}$ ($f_{RF} = f_b = |f_0 - f_R|$) at the output of the PD. At the second round-trip of the loop, the generated RF signal is sent back to the modulator. At this stage, only one single sideband modulation signal can match the RR resonance peak at frequency $f_R$ (see Fig.1b). The RR now serves as an optical bandpass filter, selecting only one sideband lobe of the modulated signal. The signal goes to the PD at the second-round trip of the loop, creating again an RF signal with frequency $f_{RF}$. After this point, the loop oscillates with an oscillation frequency at $f_{RF}$.

The main idea behind this approach is to control the frequency of the microwave signal by the wavelength spacing between the laser source and the resonance wavelength of the resonator. Since this spacing can be changed either by sweeping the wavelength of the laser or by shifting the resonance peak of the RR, this approach yields a simple tunability mechanism.

## Demonstration of the proposed tunable optoelectronic oscillator

To demonstrate the proposed operation principle, we used an integrated Si add-drop RR and external intensity modulator, photodetector and microwave circuitry (see Fig. 2). Note that all external building blocks have already been demonstrated in the silicon technology. Thus, monolithic integration of the complete OEO is technologically feasible. Nevertheless, the proposed scheme serves as a demonstrator of the principle, while providing a simple and flexible implementation, as different Si ring resonators can be tested using the same global circuit.

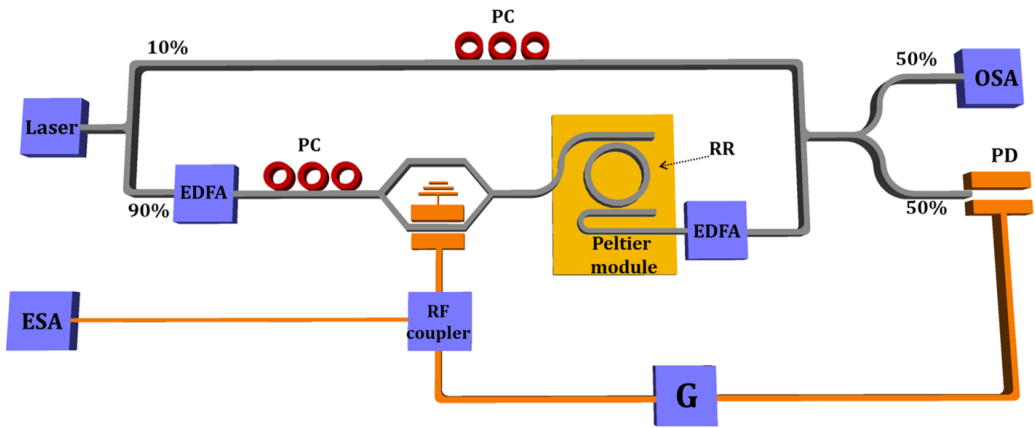

**Figure 2.** Experimental setup employed for the demonstration of the proposed tunable OEO. EDFA: Erbium doped amplifier, PC: Polarization controller, OSA: Optical spectrum analyzer, G: RF amplifier and ESA: Electric spectrum analyzer.

The Q in the add-drop ring resonator is one of the key parameters determining the performance of the proposed OEO. Higher Q yields better selectivity of the optical filter, that will determine the purity and stability of the microwave signal generated. The ring resonator was implemented on a standard SOI technology with a 220 nm thick Si thin film on top of a 3 $\mu$m buried oxide layer. We optimized the ring to operate in transverse-magnetic (TM) polarization, thereby minimizing the detrimental effect of sidewall roughness in propagation loss.

A 450 nm wide strip waveguide was chosen to ensure single-mode operation near 1.54 $\mu$m wavelength, with a resonator length L of 1 mm. In the design of the RR, adiabatic bends[19] were considered in order to reduce losses coming from the mode mismatch at the transition between straight and circular bend waveguides. A series of devices with different combinations of coupling lengths / coupling gaps were fabricated (see Methods) with the purpose to maximize the RR optical quality factor. Figure 3a shows an electron microscope image of the add-drop ring resonator. Details of the fiber-chip grating couplers and adiabatic bends are presented in Figs 3b and 3c, respectively. Figure 3d shows the measured transmission spectra (see Methods) of both through and drop ports of the RR with 300 nm coupling gap and 4.5 $\mu$m coupling length, respectively. An $FSR_\lambda$ of 640 pm was obtained accordingly, corresponding to $FSR_{fre} \approx 77$ GHz, with a RR optical quality factor $Q_{opt}$ near $8.1 \times 10^4$ (obtained by fitting the resonance peaks through a Lorentzian function).

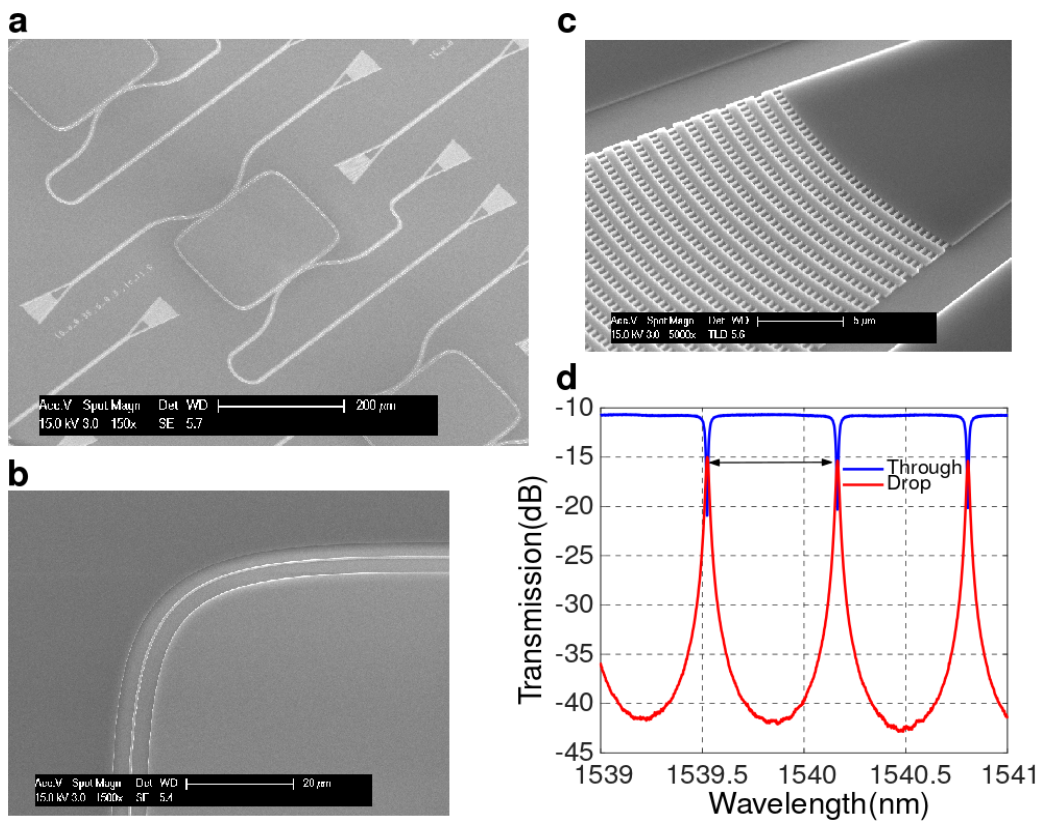

**Figure 3.** **a** Scanning electron microscope of the fabricated RR, detail of **b** Bended waveguide and **c** Grating coupler and textbfd Optical transmission of the silicon RR (coupling gap: 300 nm, coupling length: 4.5 $\mu$m).

Figure 2 shows the experimental setup used to demonstrate the proposed OEO approach. We used a 90/10 optical splitter to separate the light source coming from a CW tunable laser (Yenista TUNIS-T100S), in which 90% of the optical power was sent to an Erbium doped amplifier (EDFA) followed by the intensity modulator, the silicon RR and a second EDFA. After that, a 50/50 optical combiner was used to collect the signal from the output of a second EDFA and the optical power source signal (see Fig.2). In the experimental setup, polarization controllers (PC) were used in the upper arm of the splitter in order to match the polarization of the laser source and the signal going out from a second EDFA. At the output of the optical combiner, one arm was connected to an optical spectrum analyzer (OSA) in order to monitor the laser or resonance wavelength, while the other arm was connected to the PD. The final setup included an RF amplifier, a 90:10 RF coupler and an electrical spectrum analyzer (ESA).

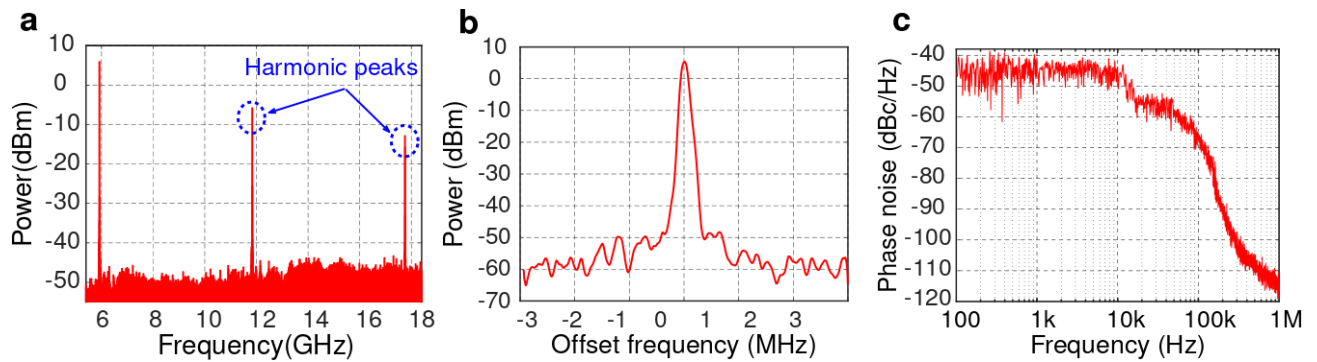

**Figure 4.** **a**) Oscillation spectrum of the generated signal based on our proposed approach, **b** the zoom-in viewed and **c** the phase noise characteristic of the created signal.

During the experiments, the resonance frequency of the RR $f_R$ was first monitored using an OSA. Then, by placing the laser wavelength (frequency $f_0$) close to an identified resonance peak, the beating between them was created. Figure 4a illustrates the electrical spectrum of the generated microwave signal within a frequency span of 13.5 GHz and with a resolution bandwidth of 200 kHz, showing an oscillation frequency at 5.9 GHz. In addition, higher-order harmonic peaks at 11.8 GHz and 17.7 GHz were also observed, caused by the nonlinearity in the OEO loop[12]. The zoomed-in view of the 5.9 GHz signal with a frequency span of 6 MHz and a resolution bandwidth of 2.2 kHz is shown in Fig. 4b, demonstrating a high signal to noise ratio of 60 dB. To evaluate the stability and the quality of the generated signal, its phase noise was measured by the automatic setup of an electrical RF analyzer (Agilent E4446A), working in the "phase noise" mode. The related result is shown in Fig. 4c, indicating a noise level of -115 dBc/Hz at 1 MHz offset frequency from the carrier. This result is comparable with the phase noise recently reported in photonic OEO implemented in silicon[12].

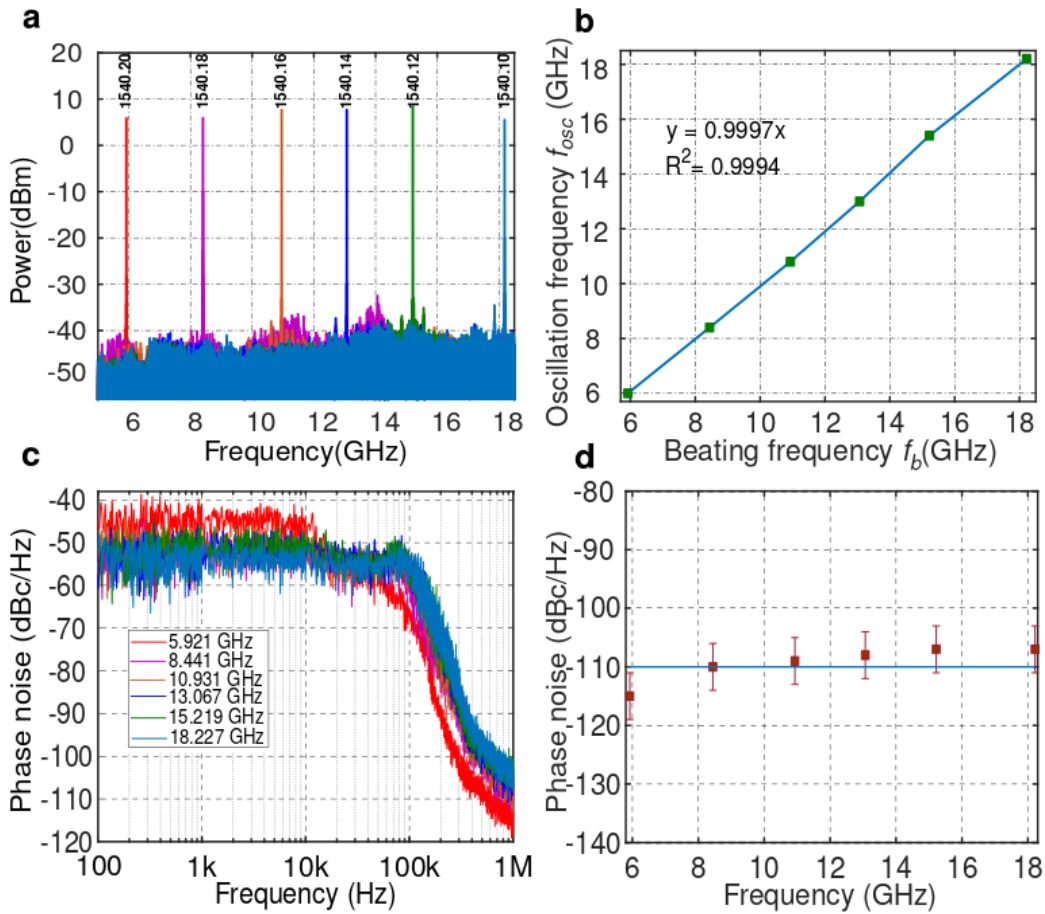

**Figure 5.** a Oscillation frequency generated with different laser wavelengths, b Plot of the oscillation frequency depending on the beating frequency. $f_b = |f_0-f_R|$, c Phase noise characteristic for differences generated signals and d Observed phase noise level at 1 MHz offset frequency from carrier.

In order to demonstrate the wide tunability of the proposed approach, we swept the laser wavelength while keeping the resonance peak unchanged. To do so, the RR sample was placed on a Peltier module to keep a constant temperature, thereby preventing resonant wavelength shifts produced by temperature changes. The RR resonance wavelength at 1541.25 nm was first observed from the OSA. Then, the laser wavelength was scanned between 1540.10 nm and 1540.20 nm. Figure 5a plots the fundamental tone of the oscillation spectrum obtained by changing the laser wavelength. These experimental results demonstrate an unprecedentedly wide frequency tunability for a Si-based OEO, ranging from 5.9 GHz to 18.2 GHz. Note that the tuning range is limited here by the bandwidth of the microwave amplifier used inside the loop.

From the corresponding frequency of the laser and resonance wavelength, we calculated the beating frequency, i.e. $f_b=|f_0-f_R|$. The evolution of the oscillation signal ($f_{osc}$) as a function of the beating frequency is shown in Fig.5b. The oscillation frequency clearly follows the beating frequency, showing a nearly perfect linear evolution with the modification of the laser frequency

separation from the RR resonance frequency (regression coefficient ≈ 0.9997). Note that in this case the RR resonance frequency is smaller than the laser one, which explains why the oscillation signal frequency increases with decreasing the laser frequency.

To evaluate the performance of the proposed tunable OEO, its phase noise characteristics have been measured for all the generated signals (see Fig.5c). Remarkably, the OEO exhibits almost constant phase noise in all the frequency spam between 5.9 GHz and 18.2 GHz. Figure 5d represents the deduced noise level at 1 MHz offset frequency from the carrier, obtained from all the curves in Fig.5c. The phase noise of the generated microwave signals are remained around -110 dBc/Hz at the offset frequency of 1 MHz, which emphasizes a key advantage of such an OEO to have a constant phase noise level with the increase in oscillation frequency[20].

## The OEO as refractive index sensor

Since the proposed OEO configuration had an oscillation frequency dependent on the refractive index environment of the RR waveguides, we have tested its characteristics for application in measuring optical index variations[21,22]. A simple approach to implement this index change was adopted by changing the sample temperature with the Peltier module, thus changing the temperature of the ring resonator, shifting its resonance wavelength. We measured this wavelength shift by wavelength scanning (see Fig. 6a) and extracted the index variation. At the same time, we monitored the variations in the oscillator frequency (see Fig. 6b). Then, as shown in Fig.6c, we could plot the oscillation frequency shift as a function of the refractive index change.

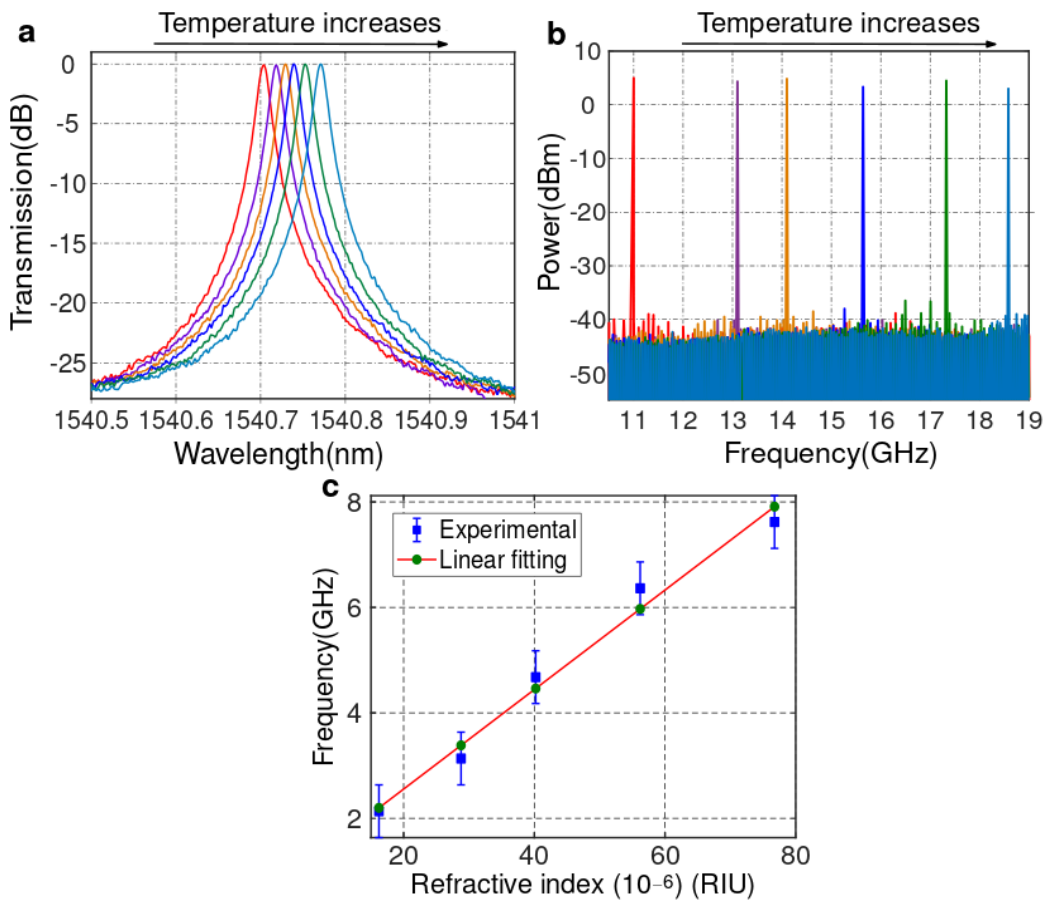

**Figure 6.** **a** Resonance wavelength, **b** Oscillation frequency simultaneously measured when temperature applied to the Peltier module increased and **c** Calculation of the oscillation change depending on refractive index variation.

In order to get a stable local temperature over the RR sample region, experiments only started 5 minutes after setting the desired temperature point in the Peltier module (in the range from 25° to 30° with 1° step size). The RR was first characterized in the optical domain. Right after the optical characterization, the closed loop OEO including the RR was measured. In the carried-out closed-loop experiment, a distributed feedback (DFB) diode laser (model 1905 LMI) operated at around 1.54 $\mu$m wavelength was used in order to provide a highly stable light source. The plot of the collected signals is shown in Figs 6a and 6b. At first glance, a fairly strong microwave frequency change can be observed between 11 GHz to 18.5 GHz. The resonance wavelength shifts towards longer wavelengths with increasing temperature. This result is in agreement with previous theoretical and experimental analyses made for SOI ring resonators[21,23]. Concurrently, the generated oscillation signal frequency also increases.

The variation in the optical index from the collected optical spectra was estimated as $\triangle n_{eff} = n_{eff} \cdot \triangle \lambda / \lambda$[24]. From a vectorial optical mode solver, the waveguide refractive index $n_{eff}$ was calculated. Considering the first detected resonance wavelength and its related oscillation frequency as the reference point, the change in refractive index was deduced accordingly, while the oscillation frequency shift from its reference value was also estimated. The deduced variation of the oscillation signal as a function of refractive index change is plotted in Fig.6c. By using a linear fitting procedure, a slope of 94350 GHz/RIU was obtained. This value is 40 times better than previously reported microwave-photonic silicon refractive index sensors using a SOI resonator device to detect frequency changes produced by cladding refractive index change[17]. In terms of limit of detection (LOD), as illustrated in the previous section, our proposed approach exhibited a stable phase noise level near -110 dBc/Hz at 1 MHz offset frequency from the carrier, allowing a system resolution of 1 MHz. From this phase noise value, a limit of detection LOD as low as $10^{-8}$ RIU can be estimated.

## Discussion

In summary, we have proposed and experimentally demonstrated a new approach for the implementation of widely tunable Si OEO. Previously reported Si OEO relied on direct conversion of intensity modulation to the microwave domain, with limited tunability, or indirect phase-modulation to intensity modulation conversion. Here, we show a direct conversion scheme providing wide tunability. In the proposed scheme, the microwave signal is created by the beating between a laser light source and a single sideband modulation signal selected by an add-drop ring resonator working as an optical bandpass filter. The microwave frequency is determined by the wavelength separation between the source and the ring resonance, providing simple tunability by sweeping the laser wavelength. Capitalizing on this concept, we demonstrate microwave signal generation between 5.9 GHz and 18.2 GHz, only limited here in the bandwidth of the employed RF amplifier. This is the widest microwave generation span reported for a Si-based OEO. Additionally, a low phase noise level of -110 dBc/Hz at 1 MHz offset frequency is achieved for all microwave frequencies, illustrating the potential of the approach for the generation of stable high oscillation frequency signals. Furthermore, we extended this approach for refractive index sensing application, harnessing high spectral resolution in the microwave domain. We have measured a sensitivity of 94350 GHz/RIU, 40 times better than state-of-the-art Si counterparts microwave photonic silicon refractive index sensor[17] and have estimated a potential limit of detection as low as $10^{-8}$ RIU for an interrogation speed of 1 MHz. We believe that the approach proposed here will expedite the development of a new generation of high-performance Si OEO with an immense potential for a plethora of applications, including, radar, wireless communications, optical signal processing, warfare systems and lab-on-a-chip biosensing.

## Methods

### Device fabrication and experimental characterization

The patterns were lithographically defined in a 100 nm ZEP-520A photoresist by using e-beam lithography. After lithography, the patterns were transferred using ICP etching with $SF_6$ and $C_4F_8$ gases. Following the waveguide fabrication, a 2 $\mu$m thick PMMA layer was deposited over the chip surface for protection.

For the optical characterization of the ring resonators a tunable laser was coupled to the input waveguide through an input grating coupler with a properly adjusted coupling angle and extracted the same way from an output grating. The grating couplers were optimized for TM polarization, yielding a fiber to fiber optical transmission of -10.5 dB at 1540 nm wavelength. A polarization controller (PC) was used to set a proper polarization at the input of the grating.

### Data availability

The data that support the findings of this study are available from the corresponding author upon reasonable request.

## Acknowledgements

This work is included in the MORSE project supported by the LaSIPS (Paris-Saclay University). The sample fabrication was performed at the Plateforme de Micro-Nano-Technologie/C2N, which was partially funded by the "Conseil Général de l'Essonne". This work was also partly supported by the French RENATECH network.

## Author contributions statement

P.T.D. and C.A.R. proposed the concept. P.T.D., E.C. and B.J. designed the devices and performed the simulations. X.L.R. and P.T.D. fabricated the devices. P.T.D., C.A.R., E.C. and B.J. performed the experimental characterizations. P.T.D., C.A.R., X.L.R., I.L., B.J. and E.C. discussed the results and wrote the manuscript. .

## Additional information

Competing Interests: The authors declare that they have no competing interests..